\documentclass[conference]{IEEEtran}
\IEEEoverridecommandlockouts
\usepackage{cite}
\usepackage{amsmath,amssymb,amsfonts}
\usepackage{algorithmic}
\usepackage{graphicx}
\usepackage{textcomp}
\usepackage{xcolor}
\def\BibTeX{{\rm B\kern-.05em{\sc i\kern-.025em b}\kern-.08em
    T\kern-.1667em\lower.7ex\hbox{E}\kern-.125emX}}
\begin{document}

\title{Assessing IT Architecture Evolution using Enriched Enterprise Architecture Models}

\author{\IEEEauthorblockN{Christophe Ponsard}
\IEEEauthorblockA{\textit{CETIC Research Centre}\\
Gosselies, Belgium \\
christophe.ponsard@cetic.be}
}

\maketitle

\begin{abstract}
Enterprise Architecture (EA) help companies to keep the evolution of their IT architecture aligned with their business evolution using a set of complementary models ranging from vision to infrastructure. In this paper, we explore how such models can be exploited to best drive this co-evolution by helping in validating different change management strategies. Considering state of the art techniques in service oriented and Cloud architecture, we propose to enrich and improve the quality of the information about the applications and software components on various qualities (e.g. performance, scalability, security, technical debt) in order to include them in the assessment process already at the Enterprise Architecture level rather than discovering them later in the change implementation phase. Our work is experimented on the LabNaf EA framework and is illustrated on a partial case study taken from an industrial project.
\end{abstract}



\section{Introduction}

Software evolvability is its ability to cope with future change in a cost effective way, especially considering changes in its environment, requirements and implementation technologies \cite{Cook00,Ciraci06}. Assessing this characteristic is highly valuable for supporting strategic decision process, especially for long-lived systems which, like most current systems, heavily rely on software for their operation.

A system or software architecture can be defined as 'the fundamental organisation of a system embodied in its components, their relationships to each other and to the environment, and the principles guiding its design and evolution' \cite{ISO42010}. The last words clearly show its key role in the evolution process. 

Over the years software have become tightly integrated in organisations to manage their information systems, starting for simple operational processing then vertically evolving towards to decision support and to strategic management. Their generalisation across enterprises also required horizontal integration to avoid functional silos and lead to the emergence of Enterprise Resource Planning and Enterprise Application Integration. Such frameworks have also evolved towards high flexibility and dynamics thanks to the use of Service Oriented Architectures \cite{Valipour07}.

\begin{figure}[h]
\centerline{\includegraphics[width=0.9\columnwidth]{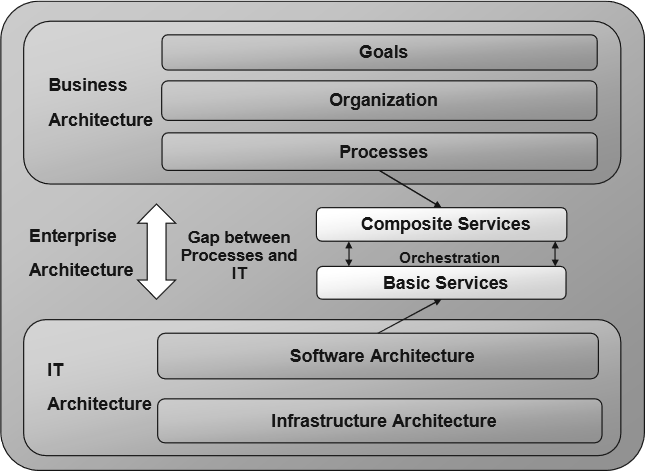}}
\caption{General structure of an Enterprise Architecture \cite{Engels08}}
\label{fig:EA}
\end{figure}

Of course organisations are also evolving in terms of goals, structure and processes, triggering the need to change the supporting software. Conversely, the adoption of new software tools enable new form of organisation. Hence, both the business and IT part of the organisation depicted in Figure \ref{fig:EA} are actually co-evolving. In order to cover all aspects in both dimensions, the notion of Enterprise Architecture (EA) is defined as the discipline for proactively and holistically leading enterprise responses to disruptive forces by identifying and analysing the execution of change toward desired business vision and outcomes \cite{Gartner14}.

Enterprise Architecture Framework (EAF) refers to any framework, process, or methodology which informs how to create and use an EA \cite{EAF18}. Although EAF can be questioned about their usefulness, adequacy or value \cite{Kotusev17}, they are widely acknowledged as providing the tooling to gather both the as-is and to-be states of the enterprise from all the involved stakeholders and to define a strategic roadmap to conduct the change. A wide variety of EAF has developed with particular advantages and disadvantages, some historical like Zachman \cite{Zachman03}, other still evolving like TOGAF \cite{TOGAF}. They are also complemented and supported by a variety of standard modelling notations such as UML \cite{UML} and BPMN \cite{BPMN2} as well as business specific way to capture, analyse or visualise enterprise data (e.g. PESTEL, Business Canvas, scorecards). Such approaches and the underlying tooling is increasingly relying on efficient modelling platform which enables many different analysis to determine the best road to take to drive the change inside the organisation. 

In the scope of our work, we focus on the evolution of the software part of the system and also stay at the macro level that is captured by enterprise architecture, i.e. high level architecture described in terms of application down to component/service-level interactions but without considering the internal implementation structure of the components. The current trends are:
\begin{itemize}
\item EA models are becoming more and more semantically rich and detailed 
\item frameworks are providing increasingly powerful analysis capabilities
\item DEVOPS tools are enabling a high level of automation of the software lifecycle and runtime monitoring making easy to harvest data about component qualities 
\end{itemize}

This combination opens new perspectives for conducting EA by considering more detailed software level information earlier. Currently, it is often fully studied only when progressing in the change implementation through specific projects and some change might reveal impractical or more costly than expected, with few alternatives at this stage. More precisely, we are interested by the following research questions:    
\begin{itemize}
\item RQ1 - What kind of analysis can be carried out from the EA model to better drive the co-evolution of the business and IT dimensions ?
\item RQ2 - What useful information should be gathered about software component and services in connection with the elaboration of business level strategies ?
\item RQ3 - What are the key issues (risks/costs) for a successful model-based approach ?
\end{itemize}

This paper reports on-going work which only provides partial answer to those questions at this stage at which we are interested by feedback from the community. Our paper is structured as follows. First section 2 gives some more context about our practical research setting in terms of framework and case study. Then section 3, 4 and 5 respectively detail some partial answers, discussed in the light of our experiment and of the work reported by others. Section 6 concludes and presents the next steps of our research roadmap.

\section{Method and Tools}

In order to explore the above research questions we used the LabNaf EAF which is a plugin of Sparx EA \cite{Labnaf,SparxEA}. We preferred it over large commercial tools like Abacus \cite{Abacus} or Alphabet \cite{Alfabet} for the following reasons:
\begin{itemize}
\item it provides a semantically consistent EA approach based on a strong underlying meta-model. This address reasoning limitations due to the variety of available and possibly overlapping frameworks (e.g. Archimate \cite{Archimate}, TOGAF \cite{TOGAF}) and modelling languages (e.g. UML, BPMN).
\item it also deals with existing standards and takes a merger approach for ensuring consistency. The ISO42010 is used for software and system architecture \cite{ISO42010}.
\item it is builds on top of standard UML extension capabilities such as stereotypes and profiles but also provides more powerful tool capabilities for analysis (queries/reports) and automation/integration (powershell) making it an interesting R\&D instruments.
\end{itemize}

\begin{figure}[h!]
\includegraphics[width=\columnwidth]{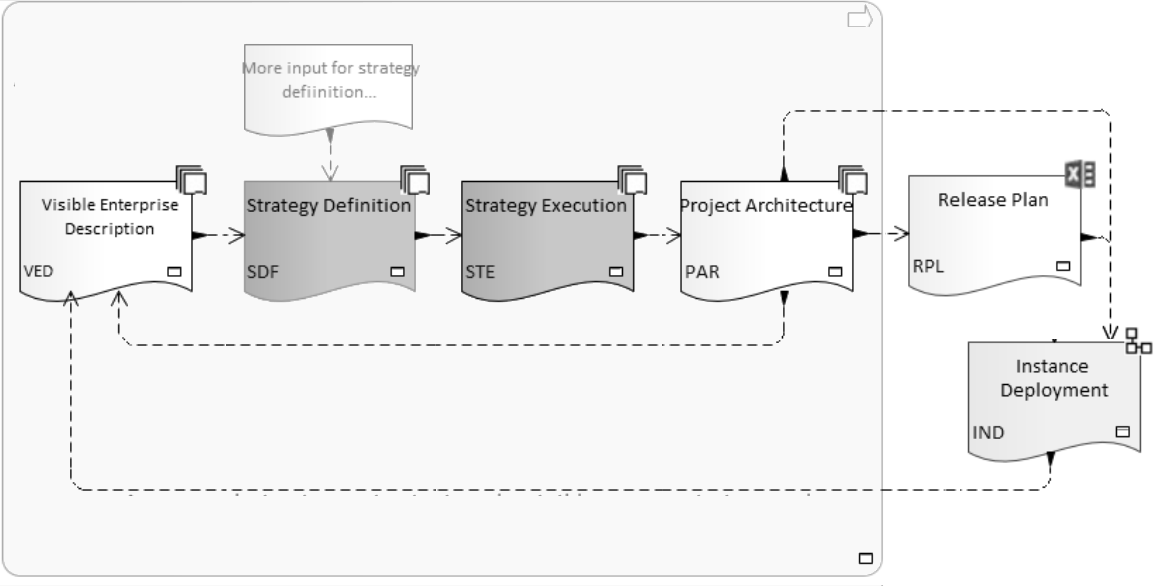}
\caption{Model of the global EA process at level 1 (including architecture)}
\label{fig:global-process}
\end{figure}

\begin{figure}[h!]
\centering
\includegraphics[width=0.9\columnwidth]{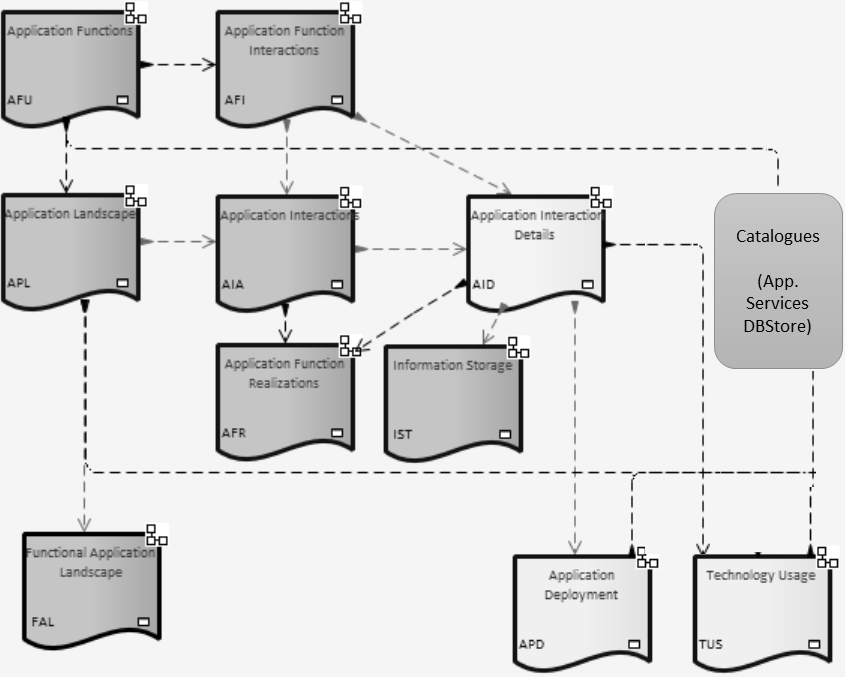}
\caption{Model of the refined architecture selection process at level 2}
\label{fig:archi-descr}
\end{figure}

The current evolution process is itself captured by the tool and described at three levels of refinement. Figure \ref{fig:global-process} shows the top level process. The visible enterprise description captures all the enterprise assets including processes, functions, information, application and infrastructure. Figure \ref{fig:archi-descr} details the typical viewpoints for capturing application and service level architectures. Note the third step of Figure \ref{fig:global-process} relates to project implementation phase and drives evolution through specific projects. It involves a number of design exploration steps that result in a decision about the solution and its impact. Once implemented it will update the enterprise description. \textbf{A key aspect is to be able to make sure relevant/high quality architecture information is taken into account in the earlier steps of strategy definition.}

On the IT side, we assume the availability of typical DEVOPS tool stack composed of continuous integration (e.g. Jenkins), software quality analysis (e.g. sonarqube), runtime monitoring (e.g. NAGIOS, ELK, Promotheus...). Those tools can generated the information for updating the EA repository.

To explore our research questions we rely on a few available evolution and migration cases studies already provided by the LabNaf framework. A local deployment of the platform is used to study the experiment on how to enrich the existing information on the repository and analyse it using relevant tools. We also looked into the literature to identify problems and solutions reported by others w.r.t. our research questions.


\section{RQ1 - Analysing Application and Service Evolution Strategies using EA Model}

A systematic review of software architecture evolution research \cite{Breivold12} identified five key themes. Without explicitly referring to  EA, the mentioned themes can be supporting by such an approach:
\begin{itemize}
\item \textit{quality consideration during software architecture design} and \textit{architectural quality evaluation} can be partly supported by the domain models relating to enterprise (see Figure \ref{fig:archi-descr}). However those need to be augmented with quality attributes such as provided by the SQUarE standard \cite{ISO25K}.
\item \textit{economic valuation}: EA support strategic decision making are typically carried out in monetary terms although it also support goal definition and different ways to assess their satisfaction.
\item \textit{architectural knowledge management}: the EA repository has a key role as centralised enterprise knowledge management tool.  
\item \textit{modelling techniques}: as discussed in Section 2.
\end{itemize}

\begin{figure}[h!]
\centering
\includegraphics[width=0.8\columnwidth]{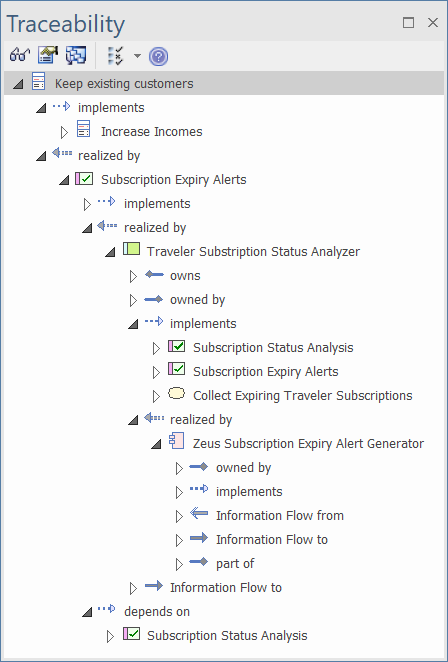}
\caption{Traceability from goal level to component level}
\label{fig:traceability}
\end{figure}

So far we have identified and experimented with some of the following analysis capabilities:
\begin{itemize}
\item traceability and impact analysis from business goal/process to application/service (and conversely) as shown in Figure \ref{fig:traceability}.
\item global ranking of component quality across the application landscape
\item identification of risk related to low quality component and critical business function
\item comparison of alternative architectures roadmaps with different plateaus (evolution milestones) to achieve the same global transformation with different tradeoffs and strategies (e.g. component replacement, component correction, component wrapping,...)
\end{itemize}

Technically the analysis could be carried out with off-the-shelf functions such as trace, impact matrix and gap analysis for the simpler ones and with query and reporting tools for more complex ones such as comparison of alternatives.

\section{RQ2 - Gathering Useful Information from Running Applications and Services}


Once the useful information is identified comes the question how to efficiently gather it. At this point we did not prototype any connector between continuous integration and operation monitoring tooling mentioned in Section 1. Beyond the technical issues, there are a several interesting issues requiring attention and already discussed in the literature \cite{Bruckmann11,Farwick12}. The later provide an interesting general check list that can drive our work:
\begin{enumerate}
\item \textit{When does the EA model have to be updated ?}\\
In our context, this cover different triggers. First, model should be updated when new release are deployed, for sure. In this case, some plateau or to-be become part of the as-is. A number of internal qualities are also available at that time if they are extracted from the continuous integration platform. Second, observing the running system will also tell us about its externally visible qualities. Those may degrade after a new deployment and can then be related to such an event but it could also be related to environmental causes (e.g. growing number of users or environment becoming hostile due to cybersecurity attacks). In this case, a difficulty is to figure out when it becomes relevant to report, it might be a progressive trend with progressive service degradation or an accumulation of punctual failures more or less impacting the business level.
\item \textit{Where does integrated data originate from (manual entry/automated update)?}\\
As we target software quality data, from the previous point two main sources can be identified: development time and run-time. A good level of automation can be expected using continuous integration platform for the former and runtime monitoring for the later. However manual collection is probably also required for specific attributes such as usability. Information collected from some manual channels such as helpdesk can also be analysed and fed into the system in a semi-automated way. More collaborative approach have also been defined based on a collaboration platform with capabilities to monitor the actual information demand and to maintain the EA model at runtime \cite{Roth13}.
\item \textit{Which part of the EA model should be updated ?}\\
In our case, the relevant part is easy to identify in the visible enterprise vision relating to application and services (see Figure \ref{fig:archi-descr}). It needs to be extended to capture the quality attributes identified in RQ1.
\item \textit{How is the quality of the resulting EA model governed, e.g. how are duplicate data entries avoided?}\\
This issue is not yet covered but discussed in RQ3.
\end{enumerate}



\section{RQ3 - Key Issues for a Successful Model-Based Approach}

\textit{Modelling semantics.} As already discussed in Section 2, high quality analysis has to rely on clear semantics of the EA modelling which is often quite light (box-and-arrow semantics, information captured in different models in an unrelated way, dealing with different granularity levels,...). We addressed this risk by selecting the LabNaf framework. At this point, the approach is quite convincing although it means some choice have been made and need to be learned, e.g. explicit restriction on Archimate. Enforcement is also not always explicit but requires to run validation tools.

\textit{Model quality.} A key point in EA is to make sure the models get updated upon changes and more generally provides a process to ensure the quality of its data. As in code a notion of technical debt can be defined and tracked. Internal consistency can be checked using validation rules against completeness, freshness or even semantic compliance of the information. On the other side, update procedure should be triggered upon all detected changes at project milestones or even on the running system (reflecting some environment evolution). When possible automated process can be used, e.g. some process can fix some well identified errors but more importantly they can be used to feed the model with information such as design time or runtime quality measurements on the software components.

\textit{Cost of model maintenance.} A major risk is that the model gets too costly to maintain over time especially when growing in size and complexity. Such cost are directly related to quality assurance and the use of automation can help in keeping it under control. Another control lever is to adapt the level of precision of the EA model in relation with the identified risks. This also motivates the selection of an approach supporting different refinement levels although it can be more complex. 


\section{Conclusion and Roadmap}

In this paper, we highlighted our on-going research to enrich EA models with information about the applications and software components on various qualities in order to better drive the global EA evolution process. We identified some research questions covering the kind of analyse required, the data collection process and the effectiveness of the approach. We also defined an experimental setting and some case studies to support our work and performed a partial literature survey.

The next steps of our research is to implement automated connectors and start gathering operations data. Going beyond our limited cases, our aim is to build a partial EA model of our research centre and to collect data about key services such as email server, website, internal applications relying on our existing monitoring and development infrastructure. Based on this, we can analyse how our current evolution roadmap is aligned with what we gather and analyse, especially related to an number of legacy yet business critical applications. We also plan to investigate more powerful data analysis tools.

\section*{Acknowledgment}

Thanks to Alain De Preter for giving early access to the LabNaf Enterprise Architecture tooling, helping in deployment and feedback on this work. 


\bibliographystyle{IEEEtran}
\bibliography{EA-EVOL}

\end{document}